\documentclass{article}

\usepackage[preprint]{neurips_2025}

\usepackage[utf8]{inputenc} 
\usepackage[T1]{fontenc}    
\usepackage{hyperref}       
\usepackage{url}            
\usepackage{booktabs}       
\usepackage{amsfonts}       
\usepackage{nicefrac}       
\usepackage{microtype}      
\usepackage{xcolor}         
\usepackage{amsmath}
\usepackage{graphicx} 
\usepackage{enumitem}
\usepackage{subcaption}

\title{Abstract Sound Fusion with Unconditional Inversion Models}

\author{%
  Jing Liu\thanks{Correspondence to Jing Liu.}, Enqi Lian, Moyao Deng\\
  Department of Music Artificial Intelligence and Music Information Technology\\
  Central Conservatory of Music\\
  Beijing, China\\
  \texttt{\{jing.liu, enqilian, 21A017\}@mail.ccom.edu.cn} \\
}

\begin{document}

\maketitle

\begin{abstract}
An abstract sound is defined as a sound that does not disclose identifiable real-world sound events to a listener. Sound fusion aims to synthesize an original sound and a reference sound to generate a novel sound that exhibits auditory features beyond mere additive superposition of the sound constituents. To achieve this fusion, we employ inversion techniques that preserve essential features of the original sample while enabling controllable synthesis. We propose novel SDE and ODE inversion models based on DPMSolver++ samplers that reverse the sampling process by configuring model outputs as constants, eliminating circular dependencies incurred by noise prediction terms. Our inversion approach requires no prompt conditioning while maintaining flexible guidance during sampling.
\end{abstract}

\section{Introduction}

Sound fusion seeks to combine an original sound with a reference sound, creating a new auditory output whose acoustic characteristics transcend simple additive mixing of the input components. In this study, sounds are classified into two primary categories: concrete sounds and abstract sounds. Concrete sounds are those that listeners can clearly associate with specific objects, scenes, events, or musical patterns that are intuitively recognizable due to their semantic content. Examples such as running water, crying children, arriving trains, and metallic percussion can be readily identified as concrete. Conversely, abstract sounds are those that do not reveal identifiable real-world sound events or, in musical contexts—particularly in contemporary, electronic, and experimental music clips—do not intuitively disclose recognizable patterns. In essence, the presence of describable, semantically meaningful patterns determines whether a sound is classified as concrete or abstract.

The classification of sounds as abstract or concrete is relative and contingent upon individual cognition and expertise. For instance, the calls of rare insects or distant whale songs may be unfamiliar to the general population and thus perceived as abstract; however, these same sounds are concrete and meaningful to entomologists and marine biologists in their respective fields of expertise.

This research specifically focuses on the fusion of abstract sounds; such fused abstract sounds demonstrate musical potential that serve as an emergent material for modern composition. We use "sound" to distinguish our focus from music, and "audio" as the carrier medium of sound. The main contributions of this work are as follows:
\begin{itemize}
\item We propose abstract sound fusion as a novel application in creative AI. 
\item We introduce an audio fusion approach that synthesizes auditory features from an original audio and a reference audio. 
\item We propose both Stochastic Differential Equation (SDE) and Ordinary Differential Equation (ODE) inversion models based on DPMSolver++ samplers that enable this audio fusion capability.
\end{itemize}

\section{Related Works}

Inversion is a technique that captures features of real samples in referenced generation, and it is usually employed in real-sample editing, a specific subcategory of generative tasks. Unlike conventional generation processes that create content ex nihilo, editing involves modifying specific attributes and characteristics of existing samples during the sampling phase. When provided with a reference, generation maintains closer adherence to users’ intentions.



Recent research in real audio editing has drawn inspiration from image editing approaches \cite{null-text, edit-friendly, pivotal_tuning, SDEdit, sto_diff_editing}, introducing new editing tasks such as substituting sound events within audio, as well as adding, removing, replacing, and refining concrete audio content \cite{AUDIT, PPAE}. This research scope expands from concrete audio to encompass abstract audio.

Most pre-existing inversion studies have aimed to reverse samples back to their initial latent codes \(x_T\), from which sampling begins. This technique originated during the GAN era \cite{manifold, GAN_inversion_survey} and has since been extended to diffusion models \cite{EDICT, BELM}. From an editing perspective, inversion addresses the challenges of unexpected changes and unwanted feature preservation that occur when employing sample-to-sample approaches.

Prior audio inversion approaches have worked with Denoising Diffusion Probabilistic Models (DDPM) \cite{DDPM} and Denoising Diffusion Implicit Models (DDIM) \cite{DDIM}, either approximating inversion through diffusion itself or by reversing the sampling process. Huberman-Spiegelglas et al.~\cite{edit-friendly} computed latent codes at subsequent timesteps directly from \(x_0\), and recent work \cite{zero-shot} extended this approach to the audio domain. Mokady et al.~\cite{null-text} worked with the DDIM sampling ODE and swapped timesteps for inversion approximation. Within the Classifier-Free Guidance (CFG) framework \cite{CFG}, the authors optimized the negative prompt during inversion while preserving the flexibility of the positive prompt. This approach fine-tuned interpolations against diffusion latents as pivots \cite{pivotal_tuning}. However, fine-tuning embeddings for each edit incurs relatively high computational costs. PPAE \cite{PPAE} subsequently utilized this approach for audio editing.

We choose the open-source Text-To-Audio (TTA) system Stable Audio Open 1.0 \cite{stable_audio_open} to achieve the best possible perceptual quality. This system uses DPMSolver++ \cite{DPMSolver++} samplers, we introduce DPMSolver++ ODE and SDE inversion models in this work. Since our approach obviates the need for prompt optimization to constrain the inversion trajectory, it preserves prompt flexibility for future creative applications. In this paper specifically, the fusion phase comprises two distinct stages, preceded by fine-tuning the model with the prompt paired with the reference audio. During the initial fusion stage, the prompt effectively guides the sampling trajectory toward reconstructing the reference audio; subsequently, following the intervention of the inverted latent of the original audio, we employ a null text prompt to complete the fusion process. 

\section{Methods}

In this section, we first introduce the notation employed throughout this paper. Next, we propose inversion models that reconstruct the original audio and describe the fine-tuning of Stable Audio Open to reconstruct reference audio. Finally, we present the audio fusion algorithm.

\subsection{Notation and Definitions}
In DDIM and DDPM, \(\alpha\) represents the portion of the signal, and \(\bar{\alpha}\) is the cumulative product of \(\alpha\), where \(\bar{\alpha}_t=\prod_{s=1}^{t}\alpha_s\). Although the notation for \(\alpha\) and \(\bar{\alpha}\) varies in the original sources, we use them consistently throughout this work.

In DPMSolver++, timestep \(t\) falls within the range \([0, s]\) for \(s>0\). Specifically, \(t\) precedes \(s\) temporally. Lu et al.~\cite{DPMSolver} transformed the diffusion ODE from the time domain to the half log-SNR domain, which is represented by \(\lambda\) and defined as \(\lambda := \log(\frac{\alpha_t}{\sigma_t})\).

The relation between \(\alpha\) and \(\sigma\) is given by \(\alpha^2_t + \sigma^2_t=1\). The model output is defined as follows: 
\begin{equation}
x^{(t)}_\theta=\frac{\boldsymbol{x}_t - \sigma_t\cdot\epsilon^{(t)}_\theta}{\alpha_t}
\label{eq:1}
\end{equation}

\(\epsilon^{(t)}_\theta\) is an abbreviation for \(\epsilon^{(t)}_\theta(\boldsymbol{x}_t,\psi)\), which denotes a network that takes the latent and prompt embeddings \(\psi\) as input conditions and outputs the predicted noise at timestep \(t\).

\(x^{(t)}_\theta\) is an abbreviation for \(x^{(t)}_\theta(\boldsymbol{x}_t, \epsilon^{(t)}_\theta(\boldsymbol{\boldsymbol{x}_t},\psi))\). It takes the latent code and predicted noise as inputs at timestep \(t\).

\subsection{Model Output as Constant}
\label{sec:constant}

DPMSolver++ introduced model output to replace the predicted noise term, which provides us with a sampling perspective: at every step toward the previous timestep, we desire the noise prediction network to predict a noise map that perfectly matches the latent such that, given both conditions, the model output exactly equals our target sample for inversion. An SDE sampler takes the following form:

\begin{equation}
\boldsymbol{x}_t=\frac{\sigma_t}{\sigma_s}e^{-(\lambda_t-\lambda_s)}\boldsymbol{x}_s+2\alpha_t\int_{\lambda_s}^{\lambda_t}e^{-2(\lambda_t-\lambda)}x_\theta^{(\lambda)}\text{d}\lambda+\sqrt{2}\sigma_t\int_{\lambda_s}^{\lambda_t}e^{-(\lambda_t-\lambda)}\text{d}\omega_\lambda
\label{eq:2}
\end{equation}

Hence, by assigning the encoded \(\boldsymbol{x}_0\) of the real sample to \(x_\theta\) in Eq.~\ref{eq:2} and removing the stochastic term, the latent at the next timestep \(\boldsymbol{x}_{t}\) is given by:

\begin{equation}
\boldsymbol{x}_t=\frac{\sigma_t}{\sigma_{t-1}}e^{\lambda_{t-1}-\lambda_t}\left(\boldsymbol{x}_{t-1}-\alpha_{t-1}(1-e^{-2(\lambda_{t-1}-\lambda_t)})\boldsymbol{x}_0\right)
\label{eq:3}
\end{equation}

Higher-order integral terms are thus eliminated from the formula, where only the schedulers are considered. Apart from the stochastic term, the drift terms of the ODE variation also differ from those of the SDE. The general ODE formula is as follows:

\begin{equation}
\boldsymbol{x}_t=\frac{\sigma_t}{\sigma_{s}}\boldsymbol{x}_{s}+\sigma_t\int_{\lambda_{s}}^{\lambda_t}e^{\lambda}x_\theta^{(\lambda)}\text{d}\lambda
\label{eq:4}
\end{equation}

Assigning \(\boldsymbol{x}_0\) to \(x_\theta\), the integral is simplified as in the SDE formula. The latent code \(\boldsymbol{x}_t\) yields:

\begin{equation}
\boldsymbol{x}_t=\frac{\sigma_t}{\sigma_{t-1}}\left(\boldsymbol{x}_{t-1}-\sigma_{t-1} (e^{\lambda_{t-1}}-e^{\lambda_t})\boldsymbol{x}_0\right)
\label{eq:5}
\end{equation}

\subsection{Audio Fusion Algorithm}
Through inversion, we obtain latents that preserve features of the original audio at different noise levels. To fuse with another audio source—namely, the reference audio—we need to capture its features as well. Therefore, we fine-tune the model to memorize the reference audio along with its corresponding prompt. We do not require generalization from this fine-tuning process; rather, our objective is to reconstruct the reference audio faithfully from the prompt.

The fusion process is a modified sampling process. The algorithm operates in two phases: in the first half of fusion, the reference audio prompt dominates to produce interpolations with implicit features of the reference audio. In the second half, we intervene the inverted latent into the sampler. 

Each term holds explicit significance at the moment of intervention. As shown in Eq.~\ref{eq:2}, the first term is where the inverted interpolation is injected, while the second term holds the interpolations guided by the reference prompt. Collectively, these two terms represent an addition of two audio sources, while the second and third terms provide flexibility for fusion.

\section{Experiments and Results}

Perceptual similarity is our primary concern for sound fusion. For general audio, especially abstract audio defined in our work, there is currently a lack of valid objective perceptual measurements due to unclear semantics, spectral complexity, and psychoacoustic influences. There exist perceptual metrics for speech \cite{JND}; however, speech occupies a narrow frequency range, whereas general audio spans the full spectrum with diverse timbral qualities, complex textures, and non-harmonic elements that speech metrics cannot adequately capture. Therefore, we employ spectrograms for intuitive visual analysis, presenting fusion results and subjective evaluation to assess fusion effects, while we assess signal quality for reconstruction.

\subsection{Original Audio Reconstruction}

We first examine the capability of our inversion models for reconstructing original audio.

\subsubsection{Necessity of Omitting Stochastic Terms}

Stochastic terms are added back during sampling because they help preserve the potential for editing. Let \(\tilde{x}_t\) be given by Eq.~\ref{eq:3}. We evaluate three different formulations for \(x_t\): (a) our proposed inversion model, (b) an approximation of prior work \cite{edit-friendly} adapted to our case, and (c) preserving the stochastic term in our inversion model.

\begin{enumerate}[label=(\alph*)]
    \item \(x_t=\tilde{x}_t\)
    \item \(x_t=x_0+\sigma_t\epsilon\)
    \item \(x_t=\tilde{x}_t-\sigma_t\sqrt{e^{2(\lambda_{t-1}-\lambda_t)}-1}\epsilon_t\)
\end{enumerate}

\begin{figure}[ht]
    \centering
    \includegraphics[width=\textwidth]{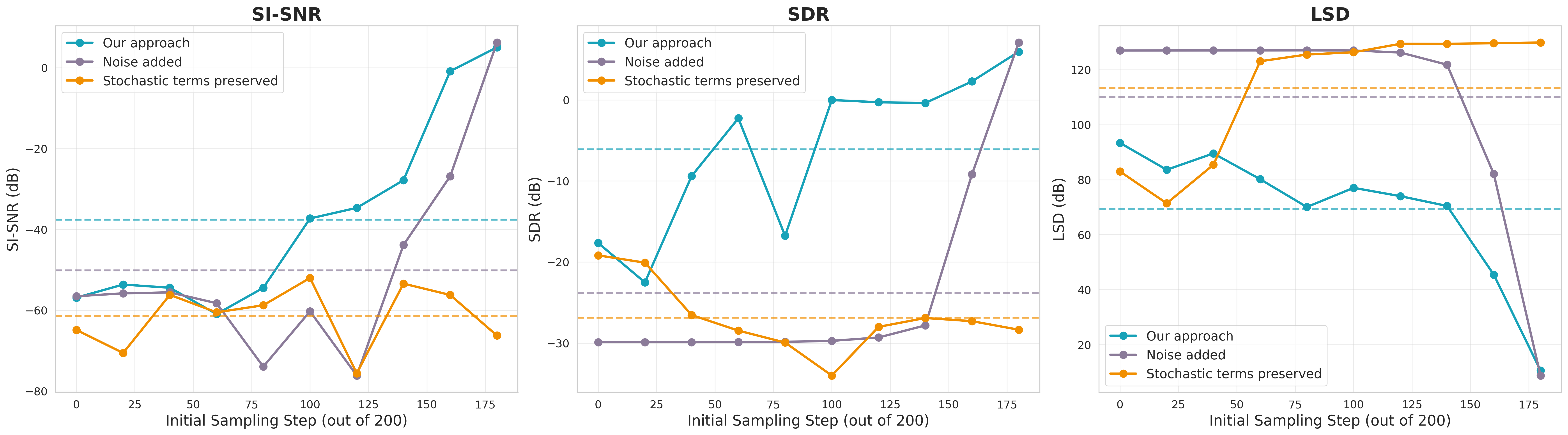}
    \caption{Audio Quality Metrics Comparison: Three Inversion Approaches. }
    \label{fig:metrics}
\end{figure}

The results illustrate that our approach achieves the best performance among the three methods, where the dashed lines represent the mean performance for each approach in Figure \ref{fig:metrics}. Any form of noise added during inversion—whether adding noise as in forward diffusion or retaining stochastic terms—results in a very noisy sampling phase. Notably, audio clips were reconstructed through Variational AutoEncoder (VAE) encoding and decoding to eliminate the influence of the VAE itself.

\subsubsection{Comparisons with Audio-to-Audio Approaches}

We compare our method with existing approaches, including:

\begin{itemize}
    \item Sampling from scratch.
    \item Image-to-image methods from the visual domain, adapted to the Stable Audio Open workflow for audio-to-audio editing.
    \item The audio-to-audio capabilities of Stable Audio 2.0.
\end{itemize}

For the open-source audio-to-audio method, we encoded the original audio through a VAE and incorporated it into randomly initialized latent representations. To ensure a fair comparison, we evaluated the SDE inversion model independently on the initial latent that deliberately avoided introducing additional adaptive editing schemes during the sampling process. We employed common knowledge prompts—those well-represented in pre-training data—as text inputs for all comparative methods. Stable Audio 2 is a commercial version\cite{stable_audio_2} whose audio-to-audio function and sampling process remain black boxes. Thus, we included it as an additional reference for comparison.

\begin{figure}[t]
    \centering
    \begin{subfigure}[t]{0.12\textwidth}
        \centering
        \includegraphics[width=\textwidth]{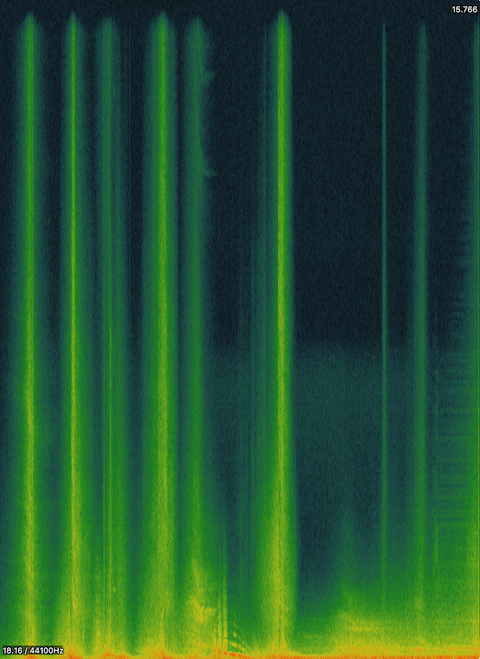}
        \caption{}
        \label{fig:2a}
    \end{subfigure}
    \begin{subfigure}[t]{0.12\textwidth}
        \centering
        \includegraphics[width=\textwidth]{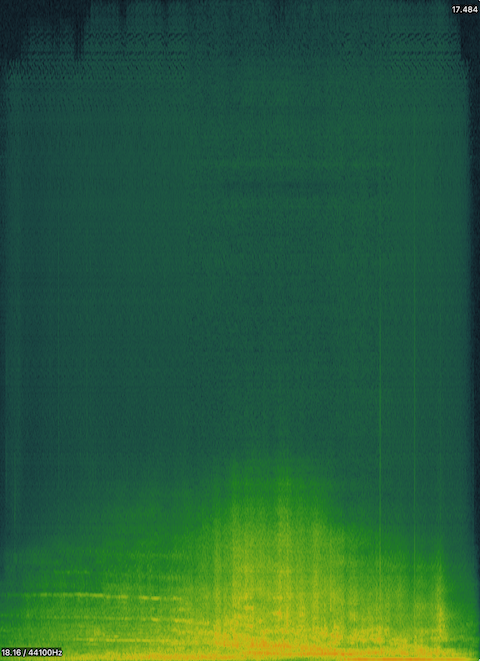}
        \caption{}
        \label{fig:2b}
    \end{subfigure}
    \begin{subfigure}[t]{0.12\textwidth}
        \centering
        \includegraphics[width=\textwidth]{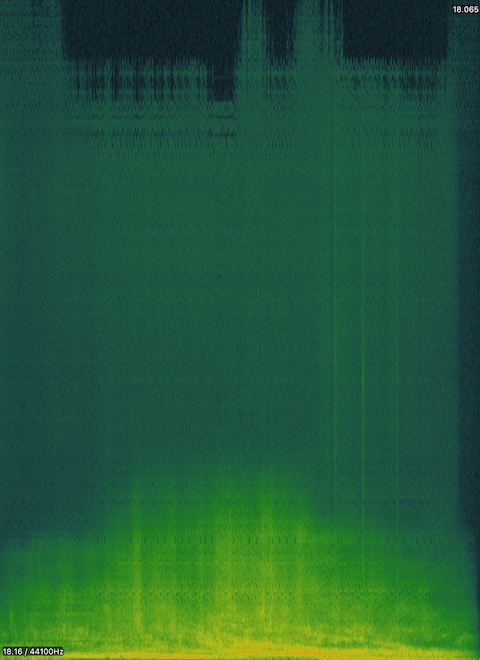}
        \caption{}
        \label{fig:2c}
    \end{subfigure}
    \begin{subfigure}[t]{0.12\textwidth}
        \centering
        \includegraphics[width=\textwidth]{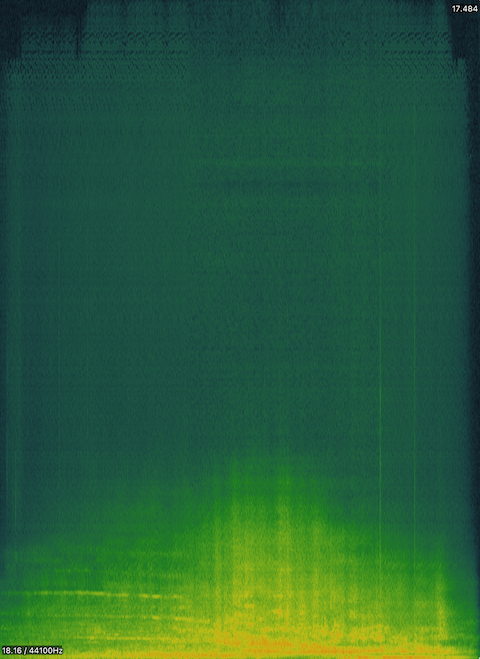}
        \caption{}
        \label{fig:2d}
    \end{subfigure}
    \begin{subfigure}[t]{0.12\textwidth}
        \centering
        \includegraphics[width=\textwidth]{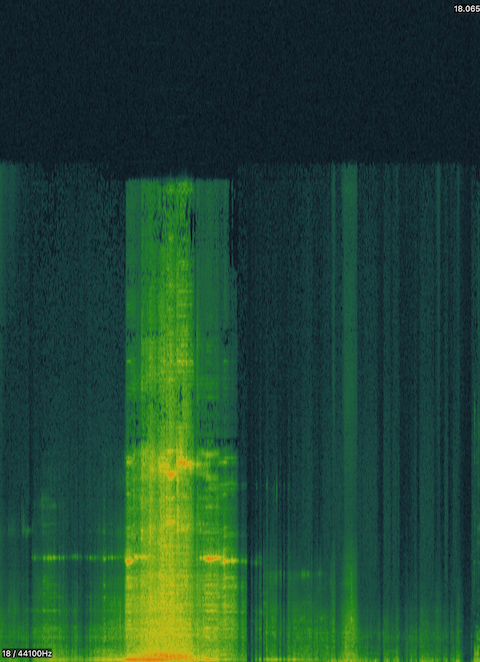}
        \caption{}
        \label{fig:2e}
    \end{subfigure}
    \caption{Comparison of audio generated from initial latents under different methods. Prompt: "An airplane takes off." (a) Original audio through VAE. (b) Sample from scratch. (c) Our inversion approach. (d) Audio-to-audio adaptation on Stable Audio Open 1.0. (e) Built-in audio-to-audio of Stable Audio 2.0.}
    \label{fig:airplane}
\end{figure}

There are subtle differences among the generated audio from the open-source model, and it is practically impossible to distinguish the original audio components across all four samples (see Figure \ref{fig:airplane}). As expected, audio generated by other methods should exhibit greater similarity to the original audio than samples generated from scratch; however, all outputs of Stable Audio Open sounded similar regardless of the original audio content. 

This phenomenon arises due to the noise control coefficient and signal strength coefficient, which make the noise guided by text prompts dominate generation during the early sampling stages. This is particularly evident in generation from scratch, where the influence of the initial latent representation is relatively weak. These observations indicate that additional controls over the sampling trajectory need to be integrated into the sampling process to achieve ideal fusion results.

\subsection{Reference Audio Reconstruction}

\begin{figure}[ht]
    \centering
    \begin{subfigure}[b]{0.12\textwidth}
        \centering
        \includegraphics[width=\textwidth]{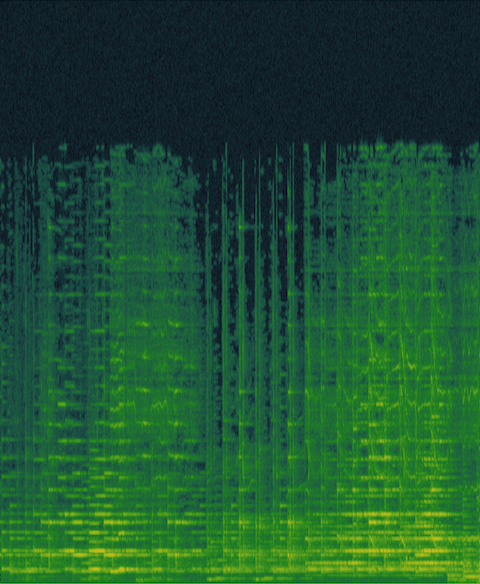}
    \end{subfigure}
    \begin{subfigure}[b]{0.12\textwidth}
        \centering
        \includegraphics[width=\textwidth]{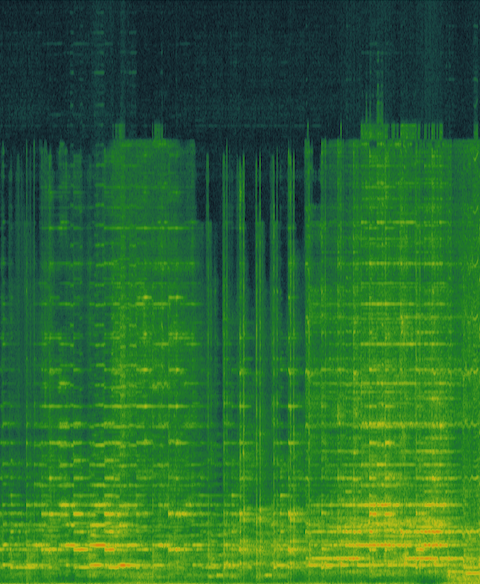}
    \end{subfigure}
    \caption{Spectrograms of reference audio (left) and reconstructed reference audio (right).}
    \label{fig:ref_recon}
\end{figure}

We fine-tuned Stable Audio Open for reference audio reconstruction. While the reconstructed audio restores the reference audio fairly completely, some detail loss and artifacts still persist. As shown in the spectrograms in Figure \ref{fig:ref_recon}, the reconstruction process generates a layer of blurry background noise, with the audio's stylistic features enhanced against this background noise. This phenomenon is most evident in the low-frequency regions. The reason for this can be found in Appendix~\ref{appx:vae}. While the audio lacks detail clarity, it is acceptable overall. Therefore, we believe that the denoising module has acquired the capability to reconstruct text descriptions into audio with reference audio characteristics.

\subsection{Sound Fusion}

We observed that inverted latent representation intervention at timesteps within a certain interval produced audio with optimal fusion results, indicating the existence of a fusion window. Within this window, as the intervention timestep decreases, the fused audio gradually transitions from resembling the reference audio to resembling the original audio in terms of auditory characteristics. Specifically, fusion represents a process of mutual balance between the reference audio and the original audio, where identifying the equilibrium point between the two within the window becomes crucial. 

We received a total of 48 valid questionnaires, with most respondents frequently participating in music activities. Music students make up the largest group with 22 participants (45.83\%); music professionals and enthusiasts each represent equal portions of the sample, with 12 participants each (25.0\%); only 4.17\% are infrequent listeners.

\begin{figure}[h!]
    \centering
    \begin{subfigure}[h]{0.45\textwidth}
        \centering
        \includegraphics[width=\textwidth]{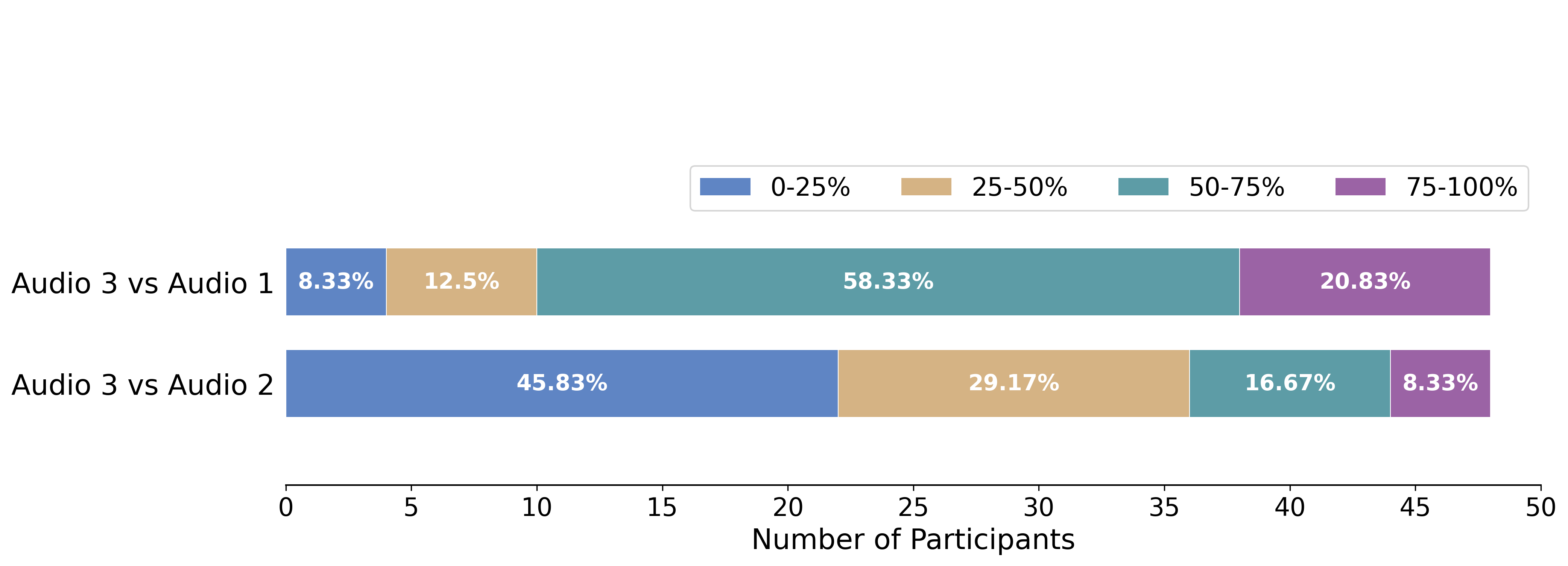}
        \caption{Set 1.}
    \end{subfigure}
    \begin{subfigure}[h]{0.45\textwidth}
        \centering
        \includegraphics[width=\textwidth]{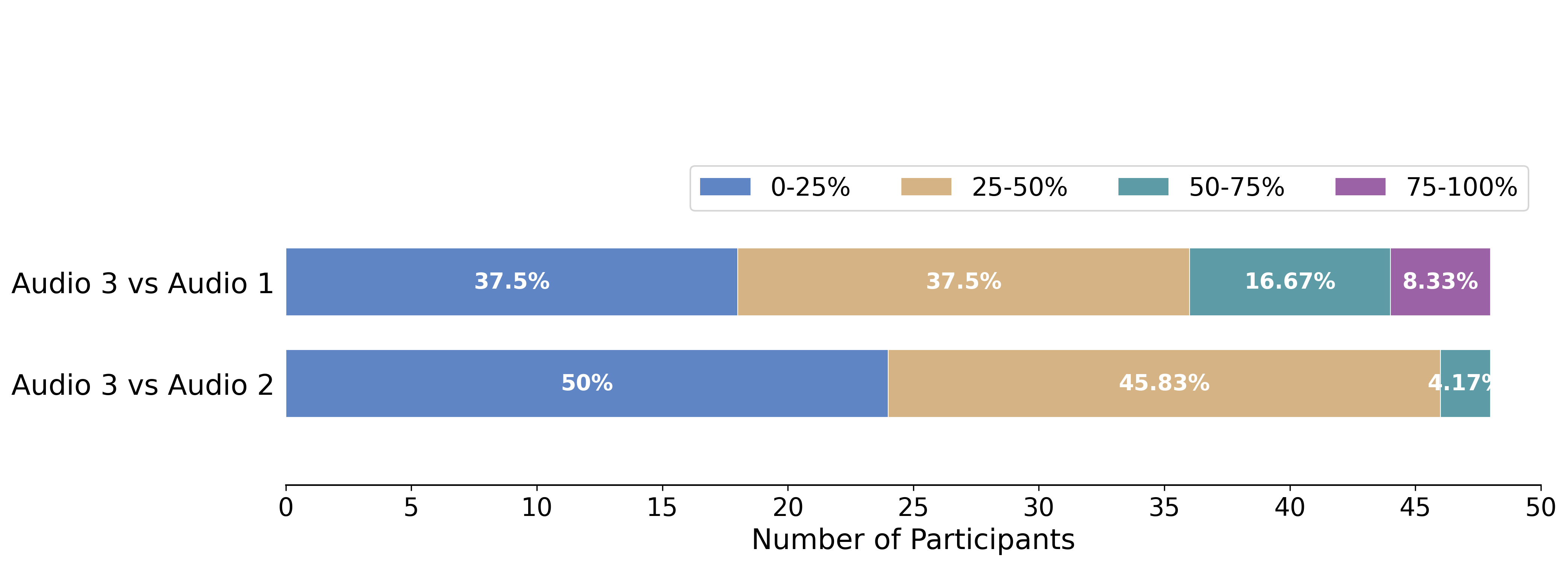}
        \caption{Set 2.}
    \end{subfigure}
    \\
    \begin{subfigure}[h]{0.45\textwidth}
        \centering
        \includegraphics[width=\textwidth]{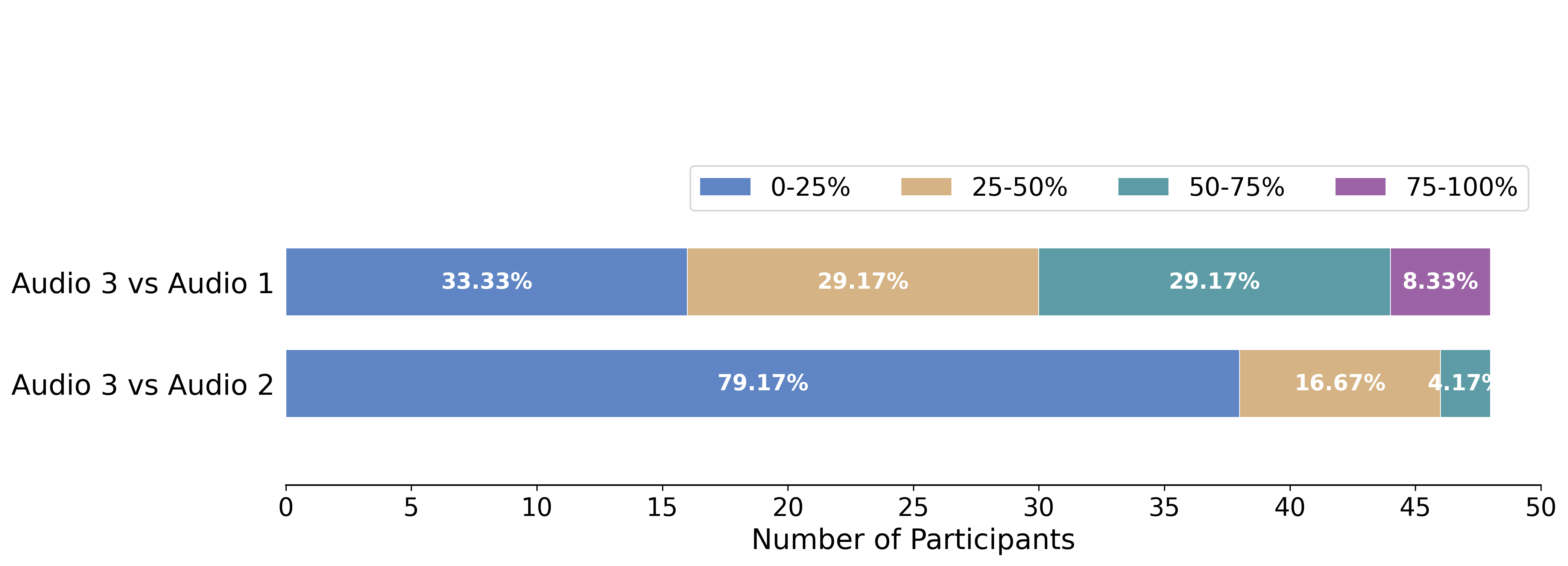}
        \caption{Set 3.}
    \end{subfigure}
    \begin{subfigure}[h]{0.45\textwidth}
        \centering
        \includegraphics[width=\textwidth]{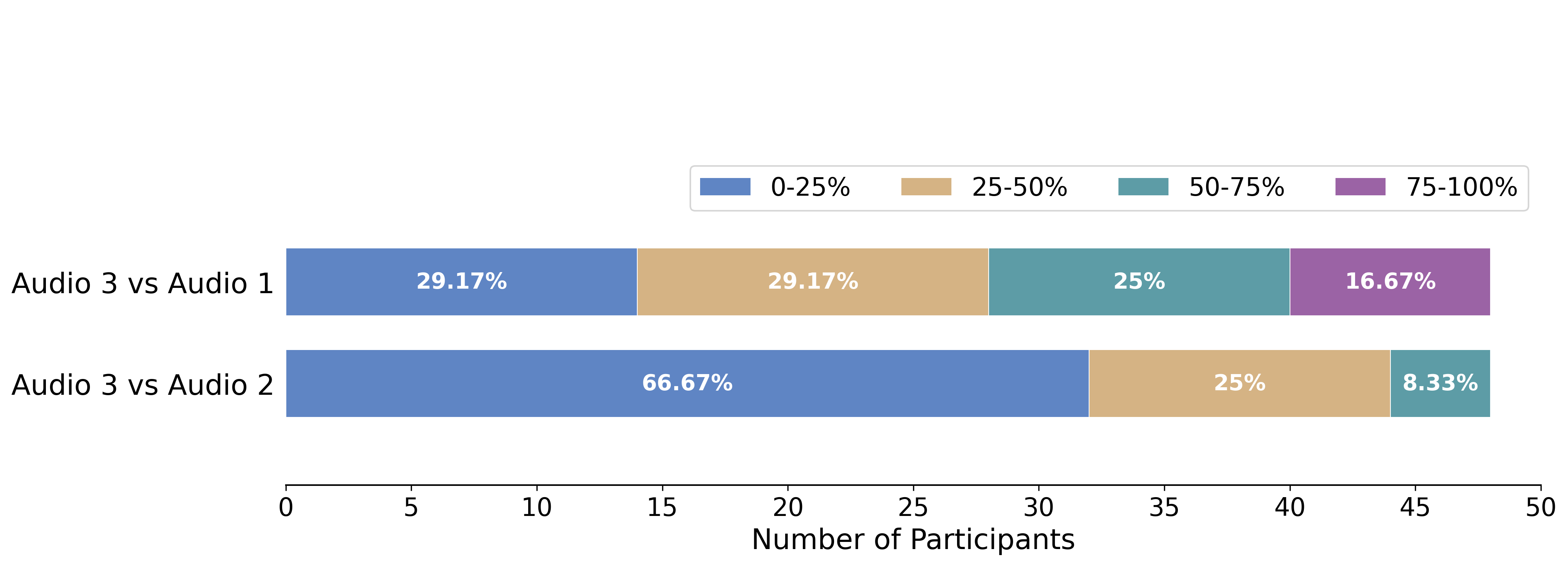}
        \caption{Set 4.}
    \end{subfigure}
    \\
    \begin{subfigure}[h]{0.45\textwidth}
        \centering
        \includegraphics[width=\textwidth]{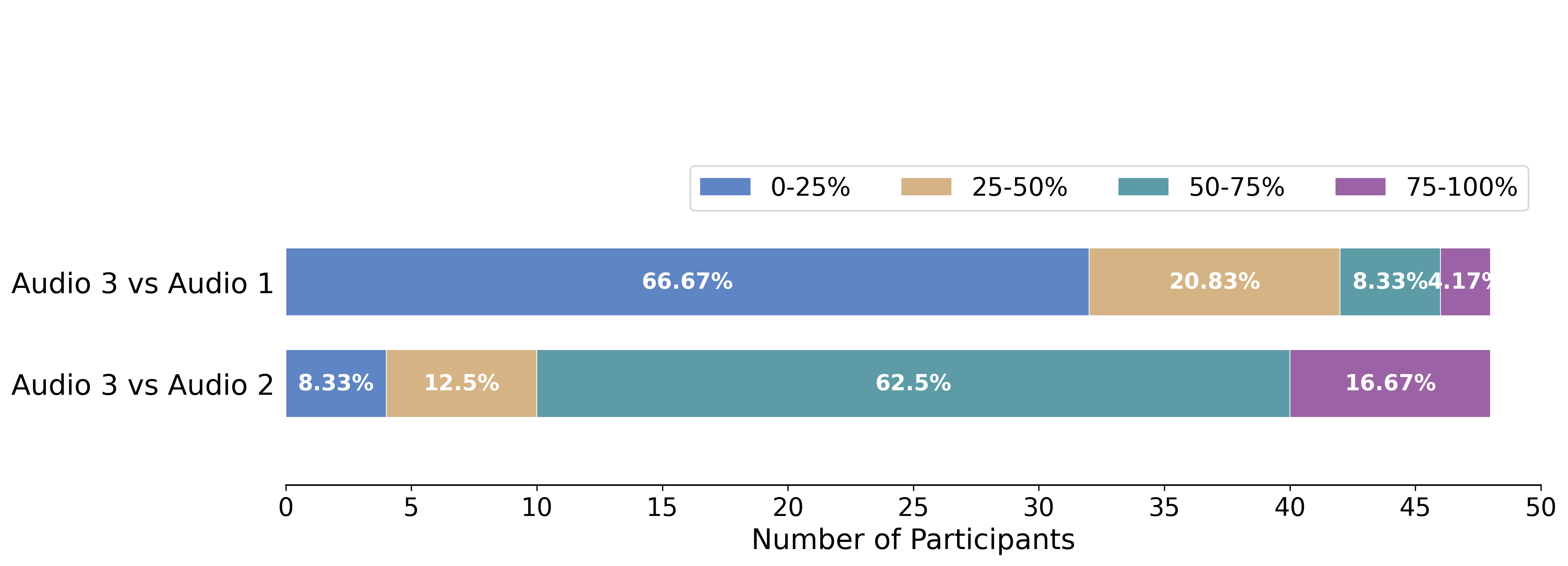}
        \caption{Set 5.}
    \end{subfigure}
    \caption{Auditory perceptual similarity evaluation results.}
    \label{fig:sub_eval}
\end{figure}

We selected fusion examples where the intervention timesteps are positioned within the fusion interval. In Figure \ref{fig:sub_eval}, "Audio 1" represents the original audio, "Audio 2" represents the reference audio, and "Audio 3" represents the fused audio; the first through fifth audio sets correspond to Set 1 through Set 5 in Figure \ref{fig:appx.set}, respectively. Overall, respondents rated the fusion effects favorably, indicating that we successfully achieved the goal of sound fusion.

\section{Conclusion}

In this paper, we propose inversion models based on DPMSolver++ with two variations: SDE and ODE. We treat the integral term as a constant, thereby avoiding the circular dependency (detailed in Appendix \ref{appx.A}) that arises from expanding the noise prediction within the sampling formula. Additionally, the deterministic formulation enables straightforward computation of the noise map at each inversion step. We explain the necessity of removing stochastic terms from the inversion model, demonstrating that approximating inversion with forward diffusion is not applicable for DPMSolver++ sampling phase.

Our experiments show that the initial latent alone is insufficient to preserve the features of inverted audio against text guidance in sampling. Therefore, it must be combined with additional control mechanisms, such as optimized prompt embeddings that guide the sampling trajectory, or intervention in the sampling phase as we propose for fusion purposes. In the latter approach, we do not necessarily need to invert all the way to the final timestep.

Inversion is employed to reconstruct the auditory features of the original audio, while fine-tuning helps preserve the features of the reference audio through prompting. An optimal interval exists during which intervention achieves effective fusion. As the timestep decreases, the auditory features of the fused audio shift from those of the reference audio to those of the original audio. Higher-order samplers provide a wider fusion interval. Due to the lack of perceptual metrics for general audio, and because our audio fusion approach does not conform to specific physical constraints, as indicated by participant feedback in Appendix~\ref{appx:sujective_analysis}, objective evaluation of our fusion results presents significant challenges. Therefore, we primarily rely on subjective auditory evaluation. The results validate the effectiveness of the sound fusion and inversion methods proposed in this work.

\medskip
{
\small
\bibliographystyle{unsrt}
\bibliography{ref}
}


\appendix

\section{Predicted Noise Introduces Circular Dependency}
\label{appx.A}

Lu et al.~\cite{DPMSolver++} proved that many high-order samplers are unstable at large guided scales, which amplified the derivatives of the noise prediction model and affect the convergence range of the ODE solver. The authors proposed replacing the noise prediction term with the model output \(x_\theta\) to achieve more stable convergence.

Previous research performed inversion on predicted noise. However, expanding the model output with the predicted noise expression results in circular dependency. To address this circular dependency issue intuitively, we substitute the general formula in Eq.~\ref{eq:2} into the second-order SDE case and target the latent at the next timestep \(r\) as below:

\begin{equation*}
\boldsymbol{x}_r=\frac{2r_1(\sigma_r^2+1)}{\alpha_t(1-e^{-2h})}\left(\boldsymbol{x}_t-\frac{\sigma_t}{\sigma_s}e^{-h}\boldsymbol{x}_s+\frac{\alpha_t(1-2r_1)(1-e^{-2h})}{2r_1}x_\theta^{(s)}+\frac{\sigma_r\alpha_t(1-e^{-2h})}{2r_1\sqrt{\sigma_r^2+1}}\epsilon^{(r)}_\theta\right)
\end{equation*}

In this example, \(\epsilon_\theta\) takes \(\boldsymbol{x}_r\) as its input, which is the variable we are solving for on the left side. This circular dependency occurs in first- and higher-order methods for both SDE and ODE samplers. Fine-tuning on latent \(\boldsymbol{x}_r\) becomes difficult because the outer scaling factor grows extremely large during the latter stages of inversion when the maximum \(\sigma\) is set to its default value. Consequently, small error accumulation leads to numerically unstable training.

\section{Derived Noise Maps}
\label{appx.B}
\setcounter{figure}{0}
\renewcommand{\thefigure}{B.\arabic{figure}}

We have inversion models for both SDE and ODE variants of DPMSolver++. We can further derive the noise map at \(t\) from Eq.~\ref{eq:1}:

\begin{equation*}
\boldsymbol{\epsilon}^{(t)}=\frac{\boldsymbol{x}_t-\alpha_t\boldsymbol{x}_0}{\sigma_t}
\label{eq:noise_map}
\end{equation*}

\begin{figure}[ht]
    \centering
    \begin{subfigure}[b]{\textwidth}
        \centering
        \begin{subfigure}[b]{0.12\textwidth}
            \centering
            \includegraphics[width=\textwidth]{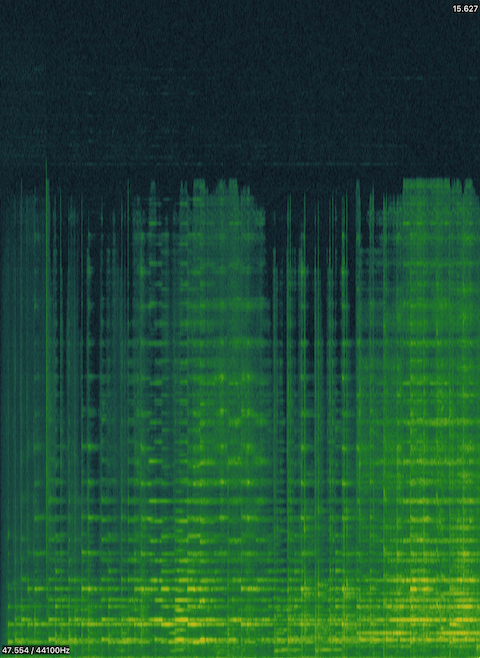}
        \end{subfigure}
        \begin{subfigure}[b]{0.12\textwidth}
            \centering
            \includegraphics[width=\textwidth]{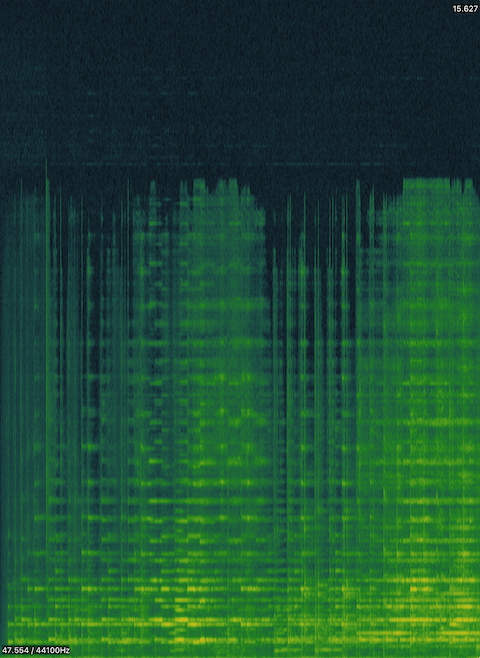}
        \end{subfigure}
        \begin{subfigure}[b]{0.12\textwidth}
            \centering
            \includegraphics[width=\textwidth]{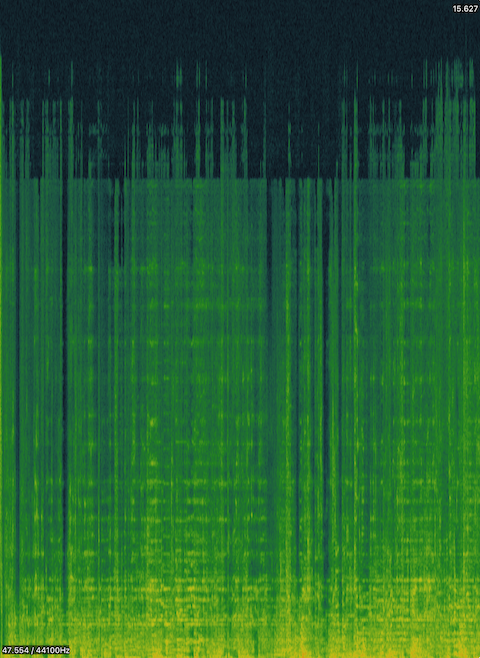}
        \end{subfigure}
        \begin{subfigure}[b]{0.12\textwidth}
            \centering
            \includegraphics[width=\textwidth]{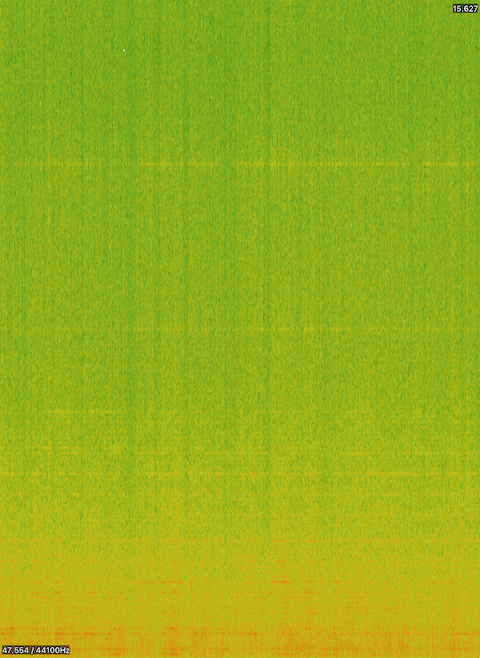}
        \end{subfigure}
        \caption{Interpolations during inversion (time steps increasing from left to right).}
        \label{fig:appxB.2a}
    \end{subfigure}
    \\
    \begin{subfigure}[b]{\textwidth}
        \centering
        \begin{subfigure}[b]{0.12\textwidth}
            \centering
            \includegraphics[width=\textwidth]{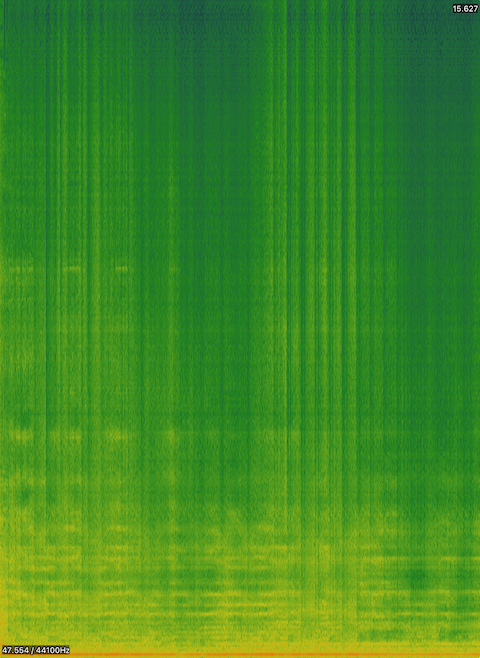}
        \end{subfigure}
        \begin{subfigure}[b]{0.12\textwidth}
            \centering
            \includegraphics[width=\textwidth]{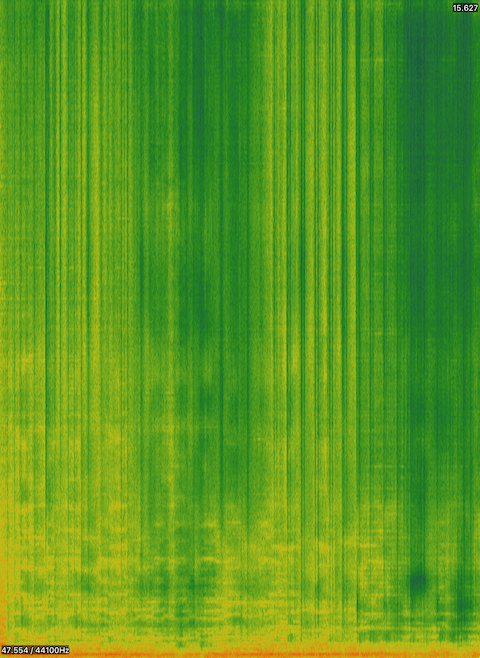}
        \end{subfigure}
        \begin{subfigure}[b]{0.12\textwidth}
            \centering
            \includegraphics[width=\textwidth]{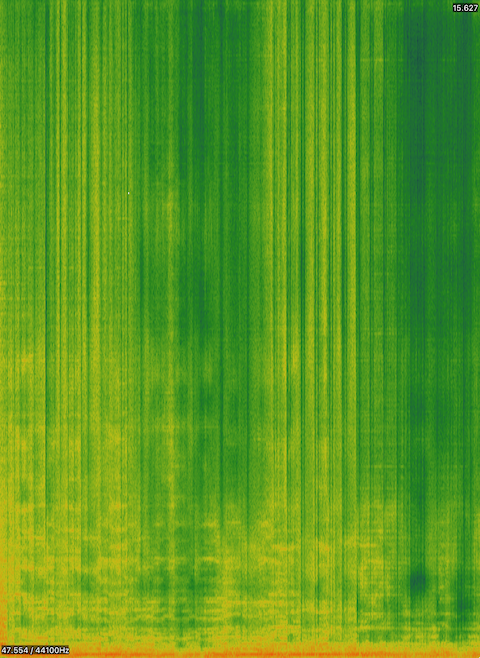}
        \end{subfigure}
        \begin{subfigure}[b]{0.12\textwidth}
            \centering
            \includegraphics[width=\textwidth]{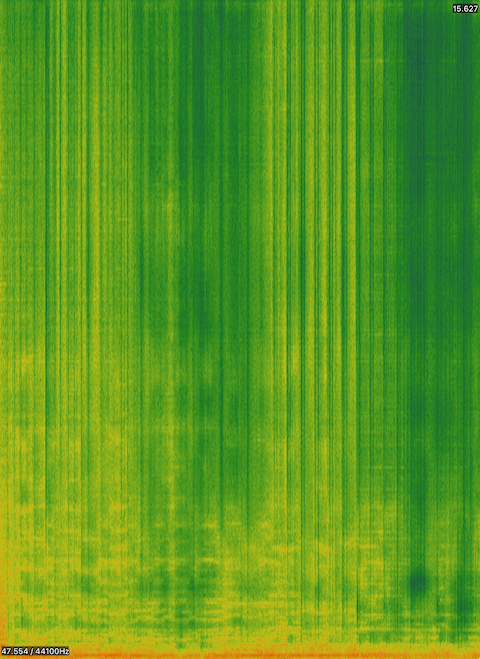}
        \end{subfigure}
        \caption{Noise maps computed from latent interpolations at the same timestep of the inversion.}
        \label{fig:appxB.2b}
    \end{subfigure} 
    \\
    \begin{subfigure}[b]{\textwidth}
        \centering
        \begin{subfigure}[b]{0.12\textwidth}
            \centering
            \includegraphics[width=\textwidth]{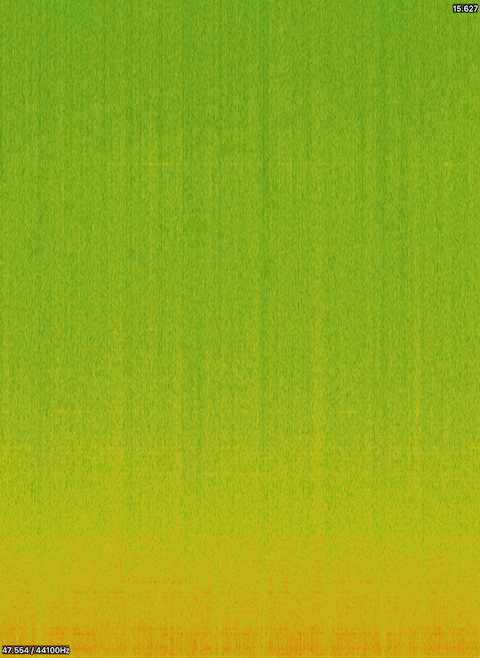}
        \end{subfigure}
        \begin{subfigure}[b]{0.12\textwidth}
            \centering
            \includegraphics[width=\textwidth]{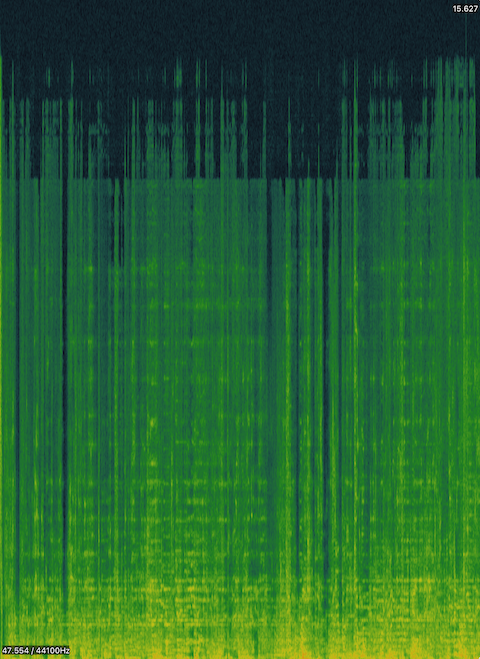}
        \end{subfigure}
        \begin{subfigure}[b]{0.12\textwidth}
            \centering
            \includegraphics[width=\textwidth]{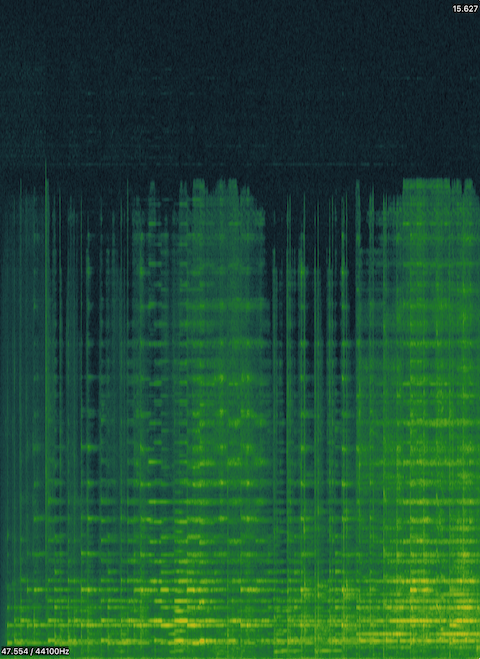}
        \end{subfigure}
        \begin{subfigure}[b]{0.12\textwidth}
            \centering
            \includegraphics[width=\textwidth]{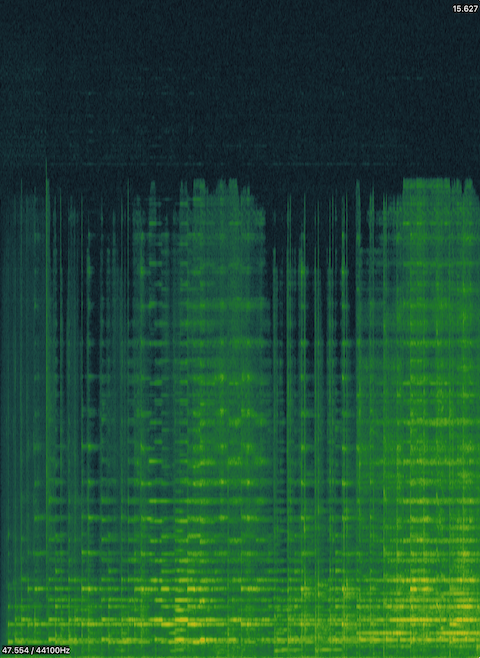}
        \end{subfigure}
        \caption{Restoration of the original audio through sampling using inverted latent interpolations and noise maps (timesteps decreasing from left to right).}
        \label{fig:appxB.2c}
    \end{subfigure} 
    \caption{Sampling from inverted latents and derived noise maps, such that the sampling trajectory coincides with the inversion trajectory.}
    \label{fig:appx.noise_maps}
\end{figure}

We demonstrate the relationships between inverted latent interpolations and derived noise maps, as well as the interpolations in the sampling process that utilize the inverted latent and its corresponding noise map at each step in Figure \ref{fig:appx.noise_maps}.

\section{Latent Intervention Across Sampler Orders}
\label{sec:orders}
\setcounter{figure}{0}
\renewcommand{\thefigure}{C.\arabic{figure}}

\begin{figure}[!htp]
    \centering
    \begin{subfigure}[h]{\textwidth}
        \centering
        \begin{subfigure}[h]{0.12\textwidth}
            \centering
            \includegraphics[width=\textwidth]{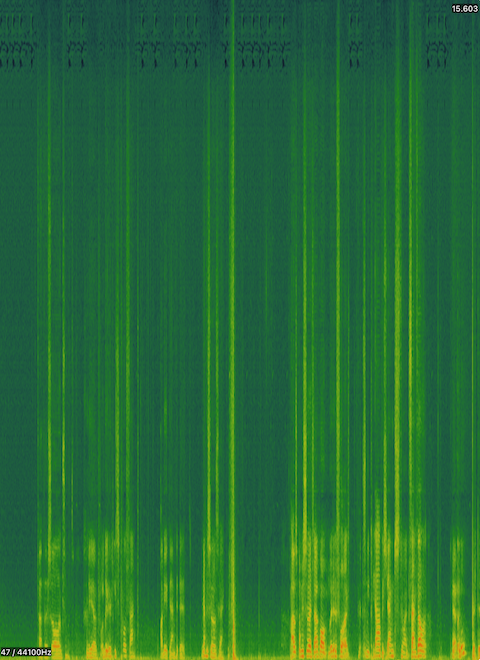}
        \end{subfigure}
        \begin{subfigure}[h]{0.12\textwidth}
            \centering
            \includegraphics[width=\textwidth]{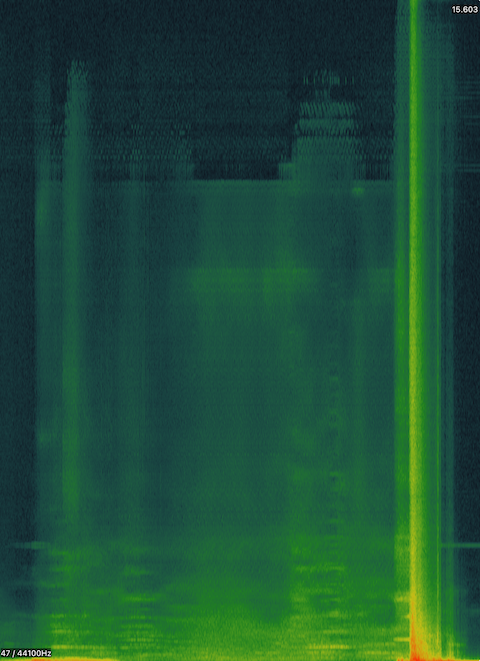}
        \end{subfigure}
        \begin{subfigure}[h]{0.12\textwidth}
            \centering
            \includegraphics[width=\textwidth]{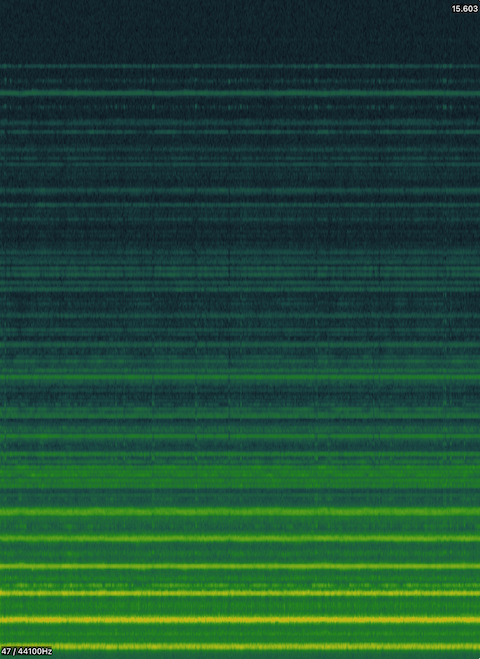}
        \end{subfigure}
        \begin{subfigure}[h]{0.12\textwidth}
            \centering
            \includegraphics[width=\textwidth]{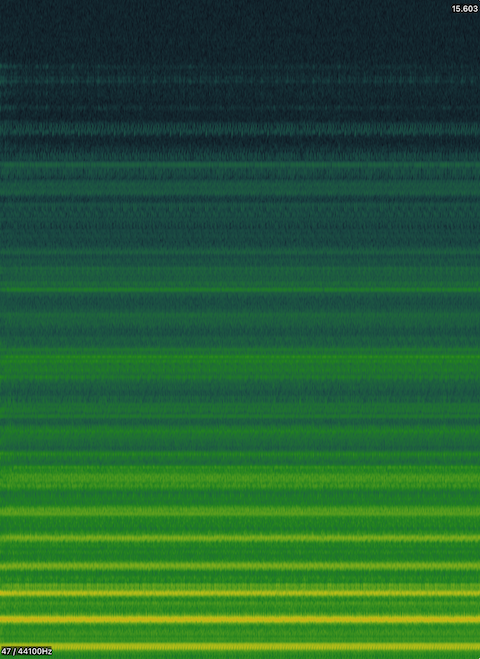}
        \end{subfigure}
        \begin{subfigure}[h]{0.12\textwidth}
            \centering
            \includegraphics[width=\textwidth]{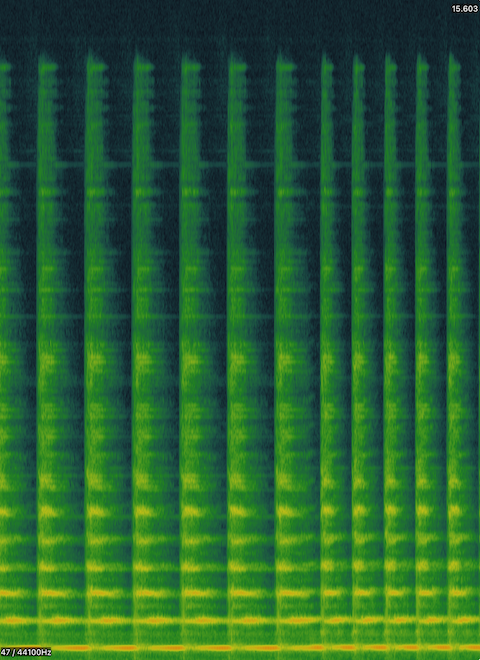}
        \end{subfigure}
        \begin{subfigure}[h]{0.12\textwidth}
            \centering
            \includegraphics[width=\textwidth]{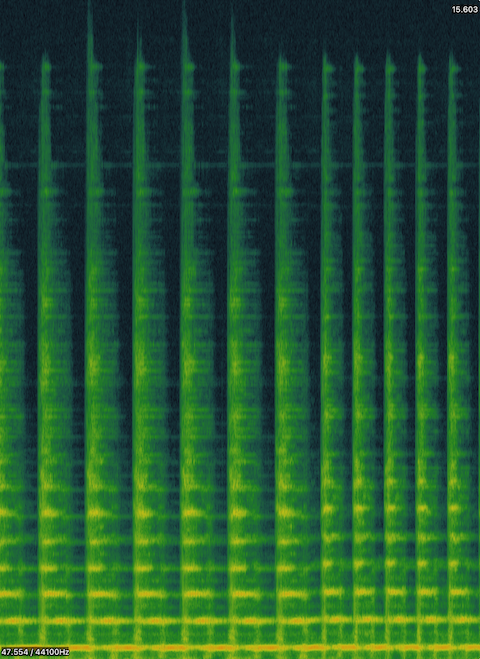}
        \end{subfigure}
        \caption{Audio reconstructed by first-order sampling.}
        \label{fig:appx.3a}
    \end{subfigure}
    \\
    \begin{subfigure}[h]{\textwidth}
        \centering
        \begin{subfigure}[h]{0.12\textwidth}
            \centering
            \includegraphics[width=\textwidth]{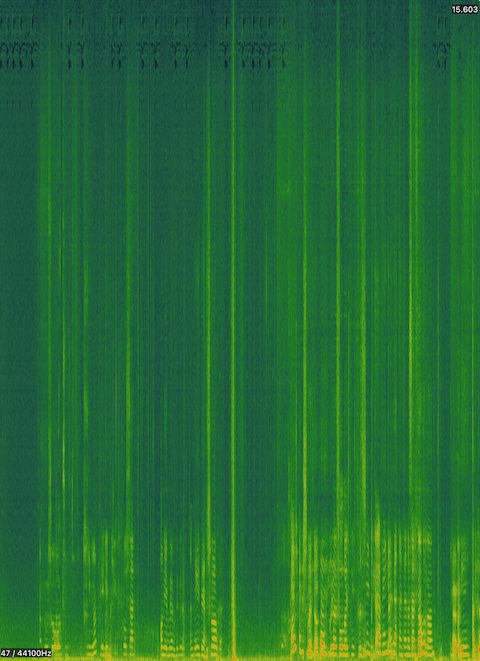}
        \end{subfigure}
        \begin{subfigure}[h]{0.12\textwidth}
            \centering
            \includegraphics[width=\textwidth]{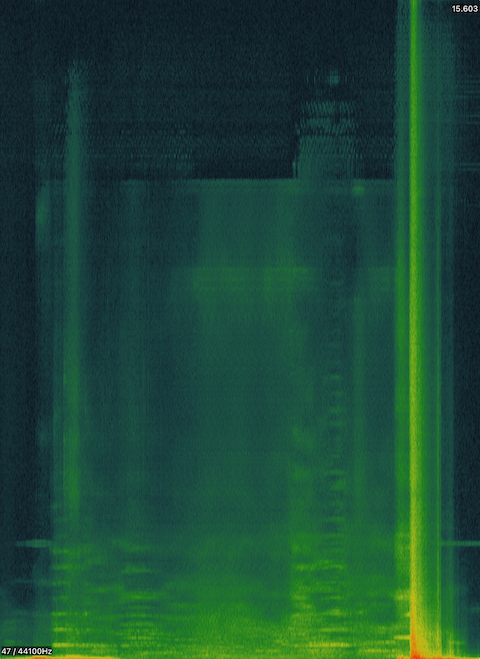}
        \end{subfigure}
        \begin{subfigure}[h]{0.12\textwidth}
            \centering
            \includegraphics[width=\textwidth]{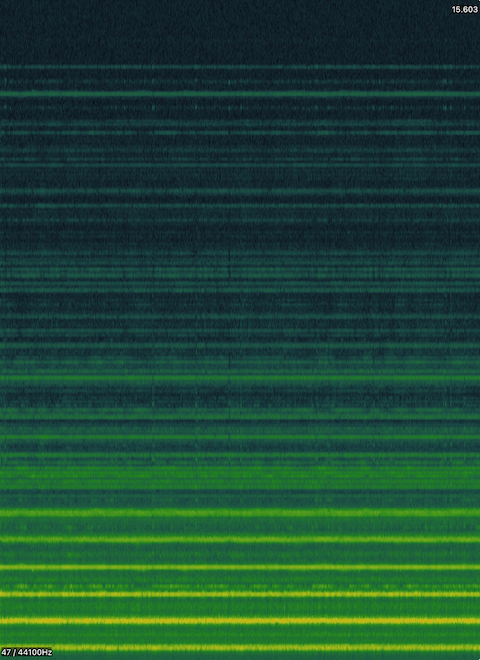}
        \end{subfigure}
        \begin{subfigure}[h]{0.12\textwidth}
            \centering
            \includegraphics[width=\textwidth]{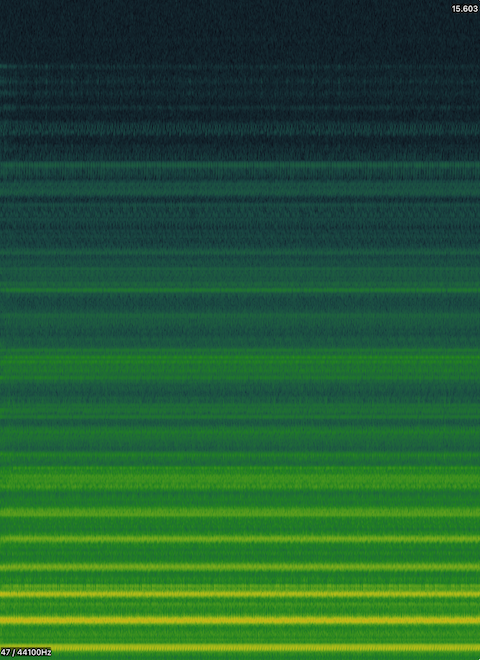}
        \end{subfigure}
        \begin{subfigure}[h]{0.12\textwidth}
            \centering
            \includegraphics[width=\textwidth]{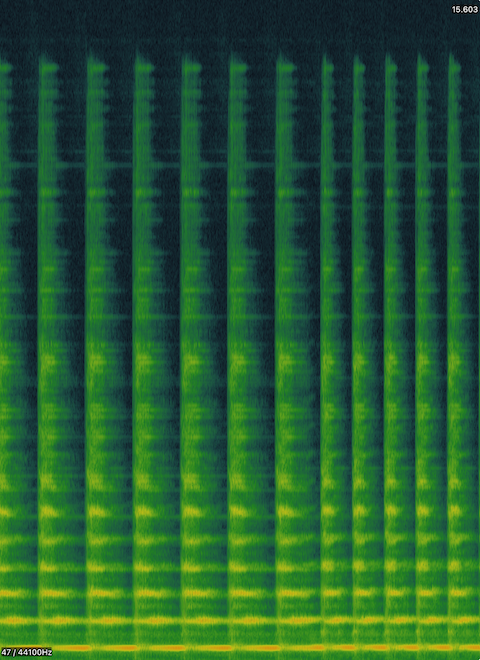}
        \end{subfigure}
        \begin{subfigure}[h]{0.12\textwidth}
            \centering
            \includegraphics[width=\textwidth]{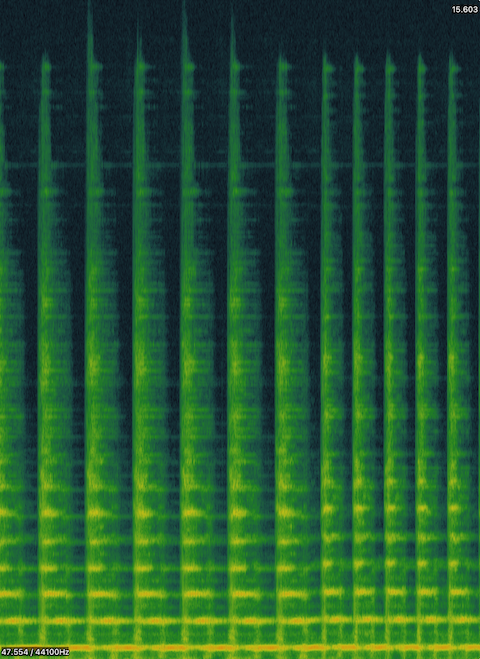}
        \end{subfigure}
        \caption{Audio reconstructed by second-order sampling.}
        \label{fig:appx.3b}
    \end{subfigure}
    \\
    \begin{subfigure}[h]{\textwidth}
        \centering
        \begin{subfigure}[h]{0.12\textwidth}
            \centering
            \includegraphics[width=\textwidth]{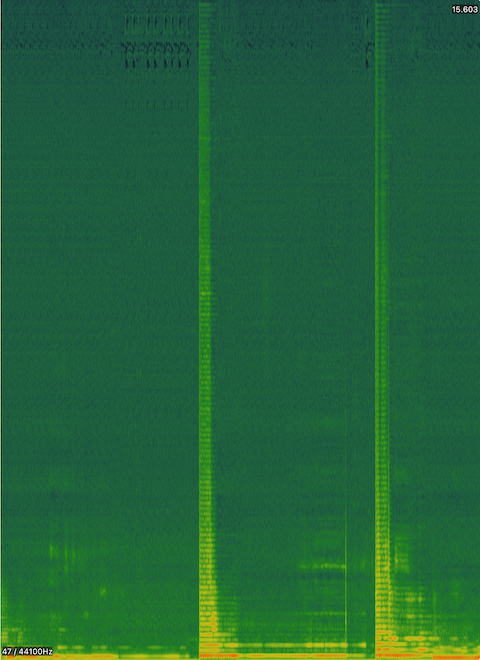}
        \end{subfigure}
        \begin{subfigure}[h]{0.12\textwidth}
            \centering
            \includegraphics[width=\textwidth]{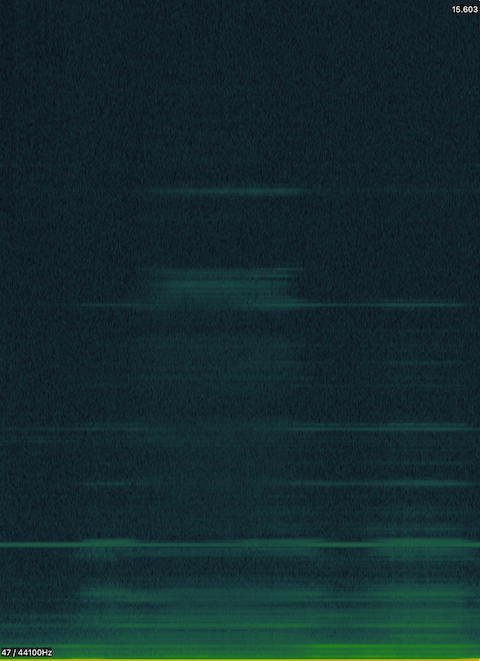}
        \end{subfigure}
        \begin{subfigure}[h]{0.12\textwidth}
            \centering
            \includegraphics[width=\textwidth]{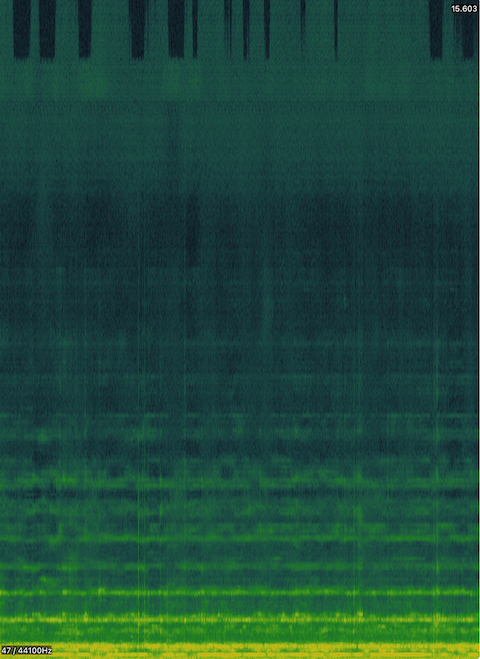}
        \end{subfigure}
        \begin{subfigure}[h]{0.12\textwidth}
            \centering
            \includegraphics[width=\textwidth]{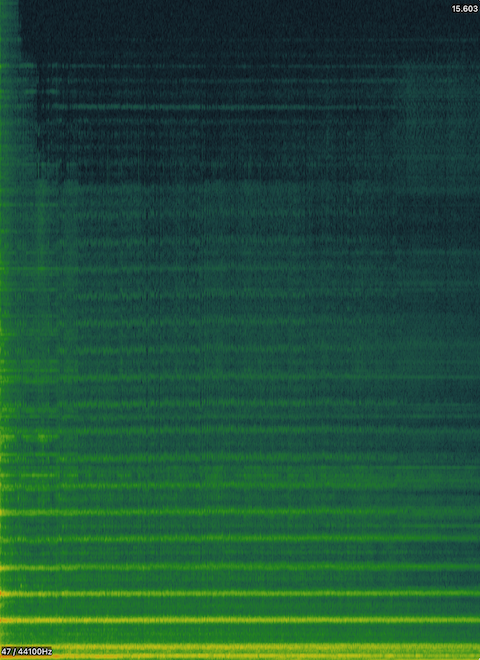}
        \end{subfigure}
        \begin{subfigure}[h]{0.12\textwidth}
            \centering
            \includegraphics[width=\textwidth]{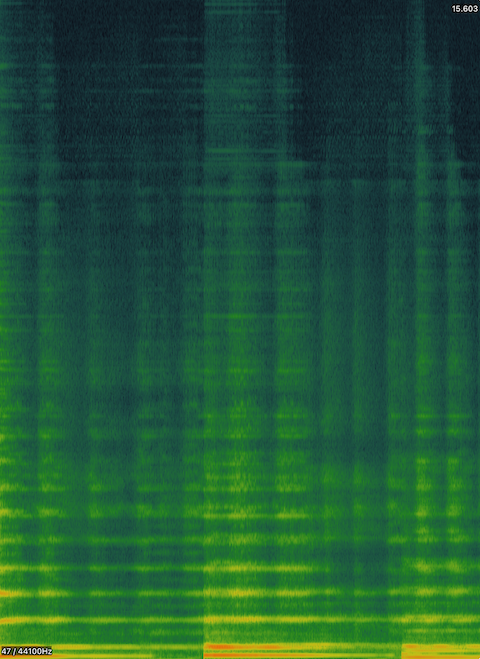}
        \end{subfigure}
        \begin{subfigure}[h]{0.12\textwidth}
            \centering
            \includegraphics[width=\textwidth]{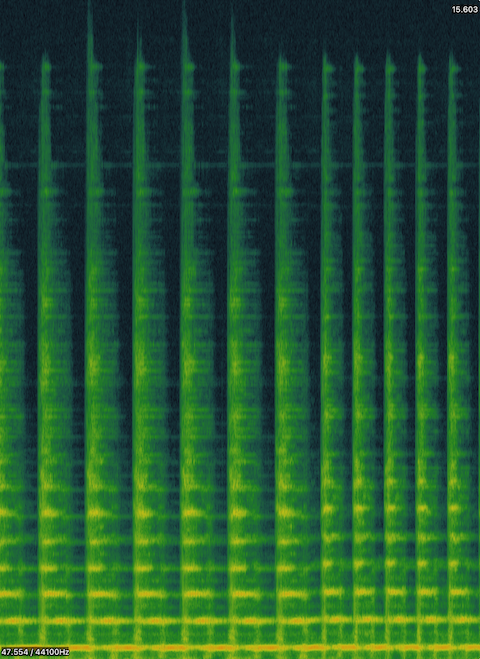}
        \end{subfigure}
        \caption{Audio reconstructed by third-order sampling.}
        \label{fig:appx.3c}
    \end{subfigure}
    \caption{SDE reconstruction after inversion with null prompt guidance. Columns 1-5: intervention at t=0, 80, 120, 140, 160. Column 6: original audio through VAE.}
    \label{fig:appx.order}
\end{figure}

Figure \ref{fig:appx.order} demonstrates reconstructed samples when intervening with inverted latent codes at timesteps 0, 80, 120, 140, and 180, compared to the sample \(\boldsymbol{x}_0\) in the last column. The three rows employ first-, second-, and third-order samplers respectively. First-order sampling is equivalent to DDIM. It is evident that features illustrated in the third-order reconstruction are relatively vague compared to the other two. Although we insert the latent at certain timesteps, higher-order methods compute with interpolations, therefore, the influence of these interpolations persists. Consequently, the intervened latents do not dominate the reconstruction to the same extent as in first- and second-order sampling. At the end of sampling, sigma, the noise scheduler, becomes very small, and the predicted noise cannot compete with the inserted latents that preserve most of the features. This indicates that initially, the prompt controls the generation process, where sigma is fairly large.

\section{Subjective Analysis on Fusion Results}
\label{appx:sujective_analysis}
\setcounter{figure}{0}
\renewcommand{\thefigure}{D.\arabic{figure}}

Fusion is achieved through music generation models, making it difficult to control granularity at the physical or acoustic level. This section presents professional electronic music analysis that reveals distinct fusion patterns across the examples.

\begin{figure}[h!]
    \centering
    \begin{subfigure}[h]{\textwidth}
        \centering
        \begin{subfigure}[h]{0.12\textwidth}
            \centering
            \includegraphics[width=\textwidth]{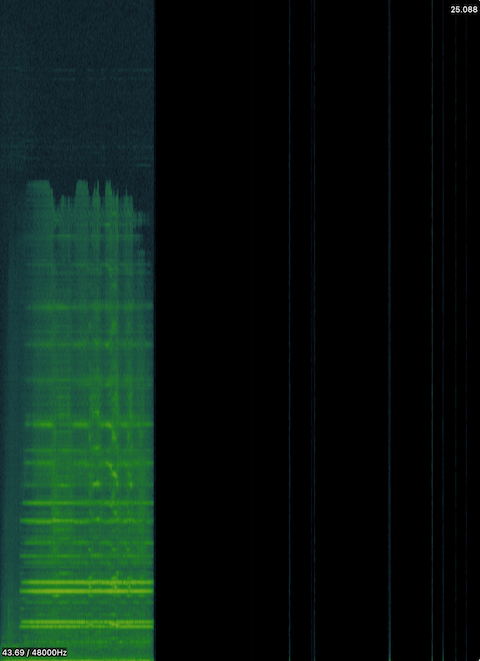}
        \end{subfigure}
        \begin{subfigure}[h]{0.12\textwidth}
            \centering
            \includegraphics[width=\textwidth]{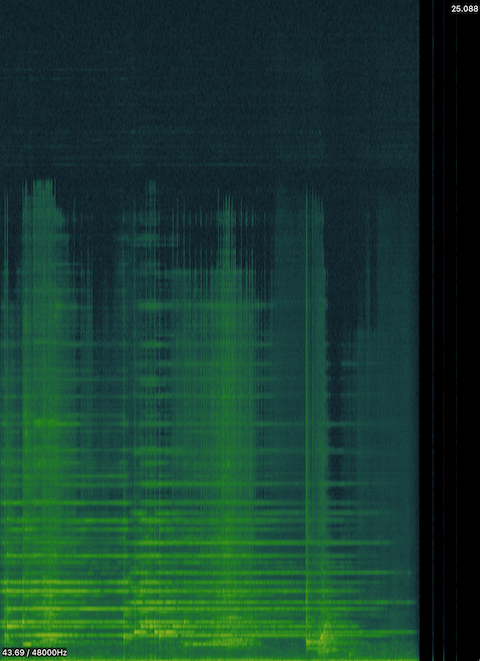}
        \end{subfigure}
        \begin{subfigure}[h]{0.12\textwidth}
            \centering
            \includegraphics[width=\textwidth]{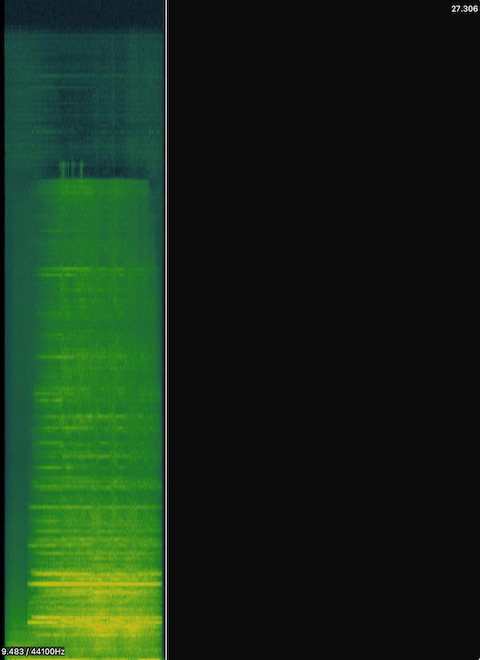}
        \end{subfigure}
        \caption{Set 1.}
        \label{fig:5a}
    \end{subfigure}
    \\
    \begin{subfigure}[h]{\textwidth}
        \centering
        \begin{subfigure}[h]{0.12\textwidth}
            \centering
            \includegraphics[width=\textwidth]{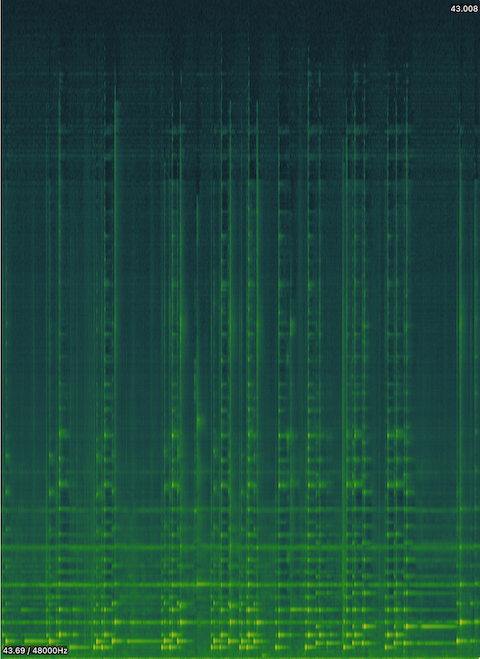}
        \end{subfigure}
        \begin{subfigure}[h]{0.12\textwidth}
            \centering
            \includegraphics[width=\textwidth]{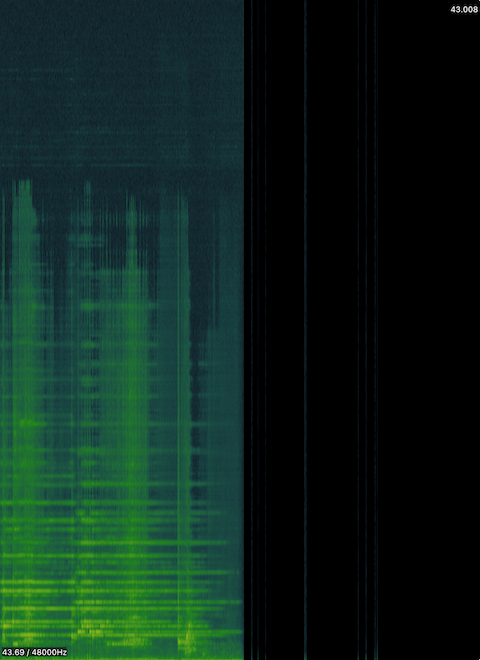}
        \end{subfigure}
        \begin{subfigure}[h]{0.12\textwidth}
            \centering
            \includegraphics[width=\textwidth]{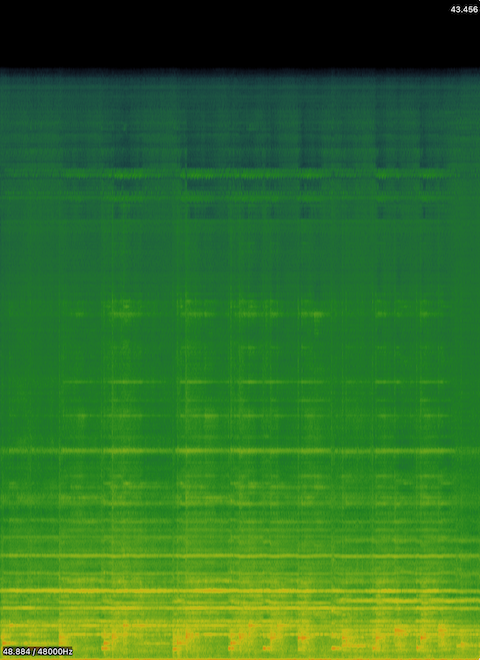}
        \end{subfigure}
        \caption{Set 2.}
        \label{fig:5b}
    \end{subfigure} 
    \\
    \begin{subfigure}[h]{\textwidth}
        \centering
        \begin{subfigure}[h]{0.12\textwidth}
            \centering
            \includegraphics[width=\textwidth]{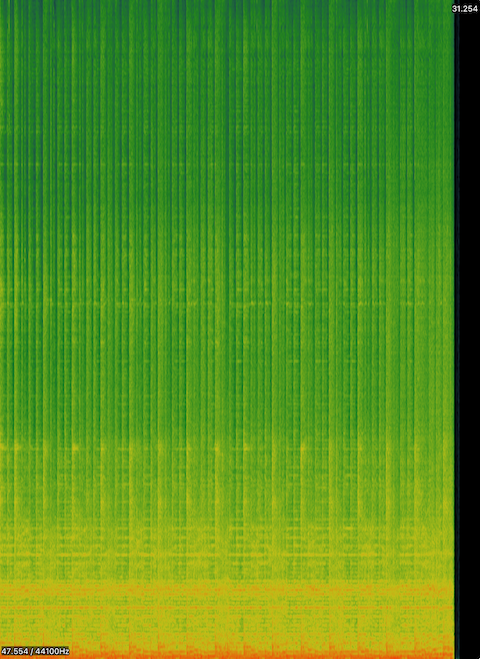}
        \end{subfigure}
        \begin{subfigure}[h]{0.12\textwidth}
            \centering
            \includegraphics[width=\textwidth]{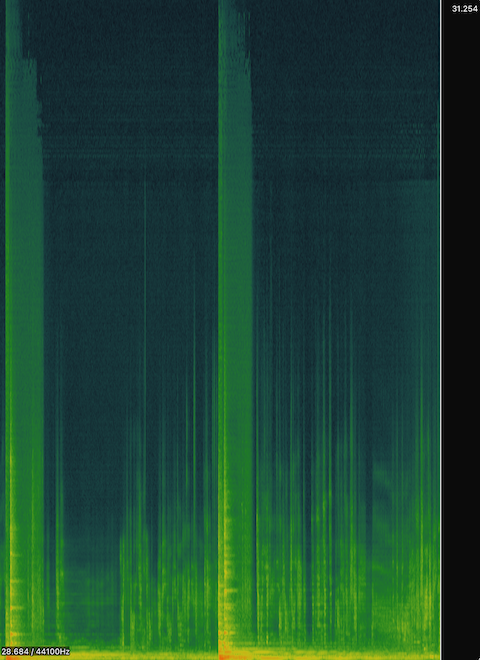}
        \end{subfigure}
        \begin{subfigure}[h]{0.12\textwidth}
            \centering
            \includegraphics[width=\textwidth]{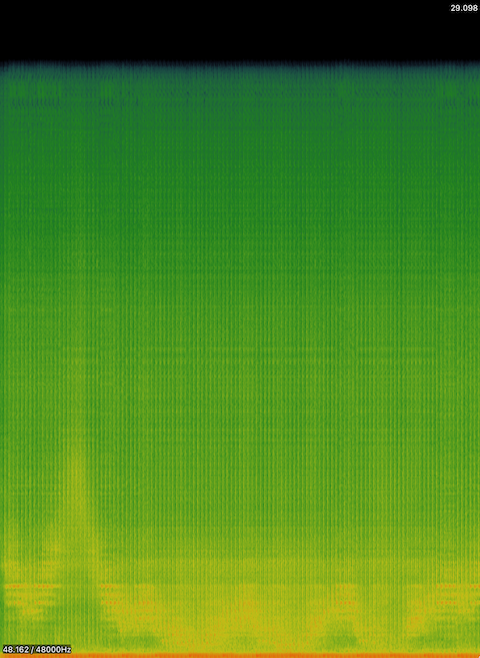}
        \end{subfigure}
        \caption{Set 3.}
        \label{fig:5c}
    \end{subfigure} 
    \\
    \begin{subfigure}[h]{\textwidth}
        \centering
        \begin{subfigure}[h]{0.12\textwidth}
            \centering
            \includegraphics[width=\textwidth]{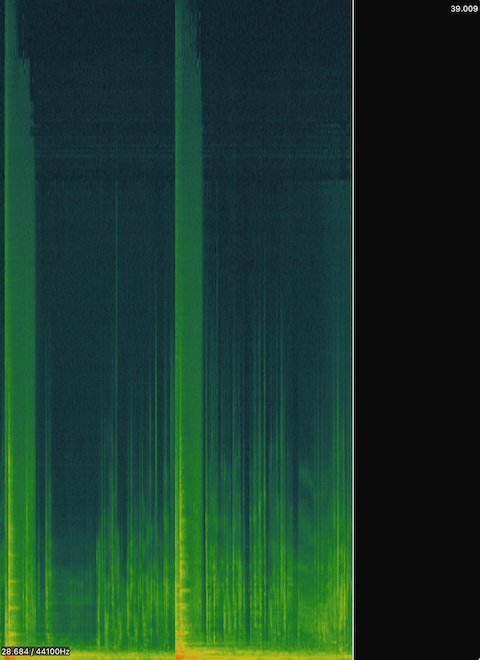}
        \end{subfigure}
        \begin{subfigure}[h]{0.12\textwidth}
            \centering
            \includegraphics[width=\textwidth]{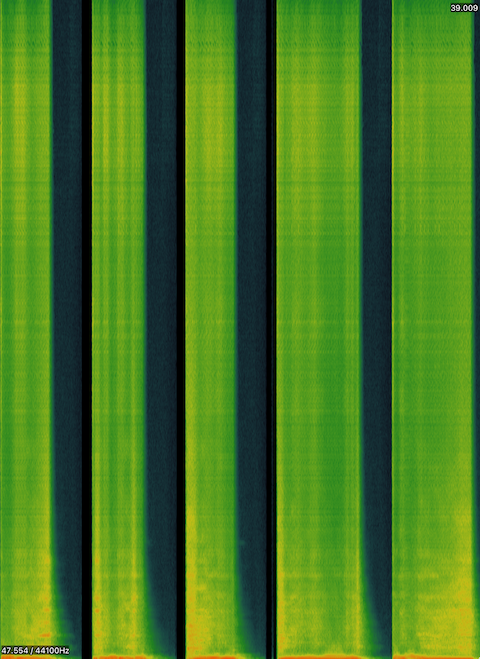}
        \end{subfigure}
        \begin{subfigure}[h]{0.12\textwidth}
            \centering
            \includegraphics[width=\textwidth]{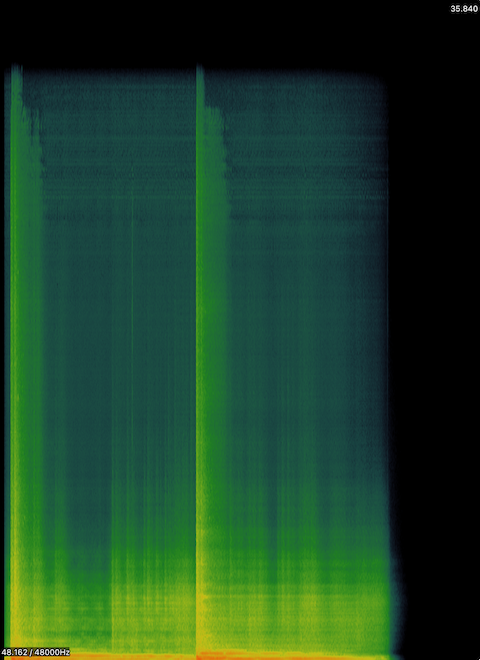}
        \end{subfigure}
        \caption{Set 4.}
        \label{fig:5d}
    \end{subfigure} 
    \\
    \begin{subfigure}[h]{\textwidth}
        \centering
        \begin{subfigure}[h]{0.12\textwidth}
            \centering
            \includegraphics[width=\textwidth]{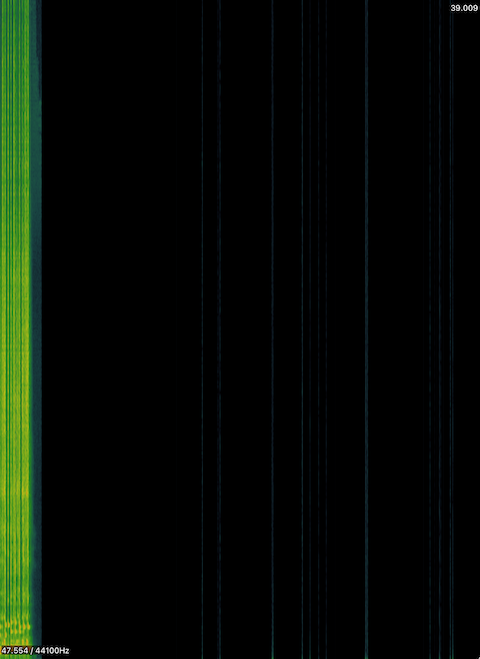}
        \end{subfigure}
        \begin{subfigure}[h]{0.12\textwidth}
            \centering
            \includegraphics[width=\textwidth]{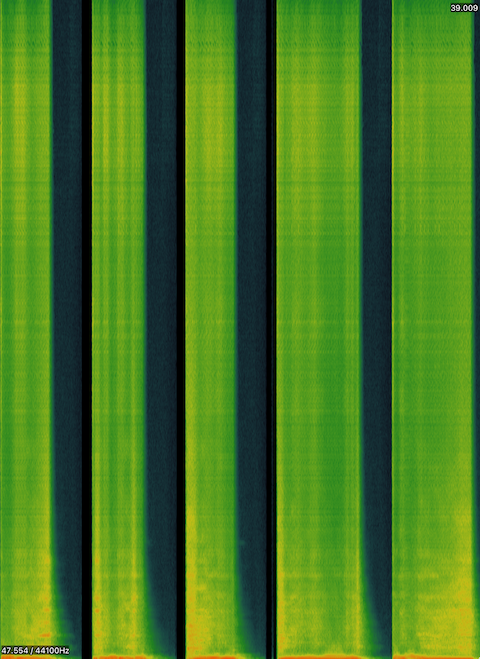}
        \end{subfigure}
        \begin{subfigure}[h]{0.12\textwidth}
            \centering
            \includegraphics[width=\textwidth]{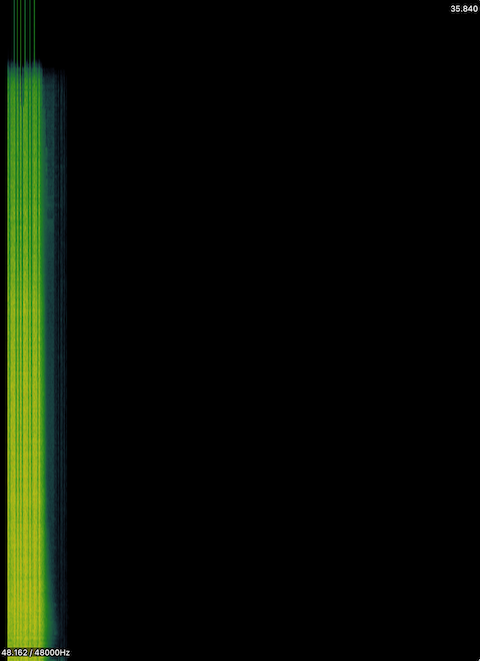}
        \end{subfigure}
        \caption{Set 5.}
        \label{fig:5e}
    \end{subfigure} 
    \caption{Audio fusion results. The first, second, and third columns represent the original audio, reference audio, and fused audio, respectively.}
    \label{fig:appx.set}
\end{figure}

In the first example of Figure~\ref{fig:appx.set}, the original audio concentrates in high frequencies, while the reference audio shows evenly distributed frequencies with prominent highs. The fused audio resembles the reference audio's frequency distribution but retains the "melodic direction" of the original audio. 

In the second example, the original audio exhibits evenly distributed frequencies with prominent highs and minimal reverb; the reference audio has obvious melodic characteristics with concentrated high tones and some reverb; the fused audio has significant reverb and combines melodies from both sources. 

For the third example, the original audio has obvious rhythm, mainly percussion-based with repetitive melody; the reference audio has a wide melodic range with almost no rhythm and sustained low frequencies; the fused audio combines the rhythmic nature of the original audio with the sustained and repetitive characteristics of the reference audio's low frequencies. 

In the fourth example, the original audio has a wide melodic range; the reference audio is essentially pure noise; the fused audio basically repeats the original audio's melody but incorporates substantial low-frequency noise from the reference.

In the fifth example, the original audio is primarily rhythmic with percussion with layered melody; the reference audio is pure noise with spatial characteristics; the fused audio combines the rhythm and melody of the original with the spatial characteristics of the reference, amplifying these effects.

\section{VAE Blurs Acoustic Details}
\label{appx:vae}
\setcounter{figure}{0}
\renewcommand{\thefigure}{E.\arabic{figure}}

\begin{figure}[ht]
    \centering
    \includegraphics[width=0.35\columnwidth]{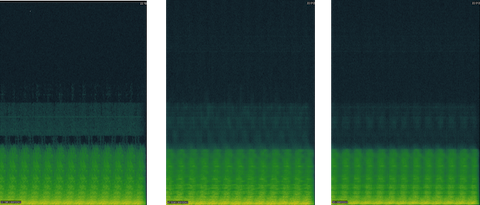}
    \caption{Spectrogram of the original audio (left), spectrogram of audio reconstructed through VAE encoding-decoding (center), and spectrogram of audio reconstructed from inverted \(x_t\) at step 100/ 200 using null prompt guidance (right).}
    \label{fig:appx.vae} 
\end{figure}

Initially, the audio is encoded by the VAE to derive \(x_0\). Figure \ref{fig:appx.vae} presents a comparative analysis. While GANs typically produce high-definition samples, VAEs generate comparatively smoother outputs. Constrained by these limitations, VAE encoding-decoding procedures result in perceptible attenuation of acoustic details, manifested in the spectrogram as indistinct boundaries and homogenized frequency band stratification. This phenomenon derives from a fundamental VAE property: latent space continuity. To enhance generation fluidity and minimize artifacts and distortion, proximate points within the VAE latent space decode to similar outputs, though these points may not correspond to actual encoded inputs. The right-side representation, when compared to the center, exhibits additional loss of features in upper frequency bands that possess low energy. Nevertheless, the VAE preserves the macro-structure, with pronounced periodic characteristics remaining intact. The left-center comparison serves to demonstrate inherent model constraints, isolating these from potential effects attributable to the inversion models themselves. Consequently, evaluations in this paper compare reconstructions through VAE encoding-decoding with original samples, rather than comparing with the original samples directly.

\end{document}